\documentclass[showpacs,preprint,amsmath,amssymb]{revtex4-1}

\usepackage{graphicx}
\usepackage{dcolumn}
\usepackage{bm}
\usepackage{color}

\newcommand{\subm}[1]{_{\mathrm {#1}}}

\begin{document}

\preprint{APS/123-QED}

\title{Band-selective Plasmonic Polaron in Thermoelectric Semimetal Ta$_2$PdSe$_6$ with ultra-high power factor} 

\author{Daiki~Ootsuki$^1$}
\email{ootsuki.daiki@okayama-u.ac.jp}
\author{Akitoshi~Nakano$^2$}
\author{Urara~Maruoka$^2$}
\author{Takumi~Hasegawa$^3$}
\author{Masashi~Arita$^4$}
\author{Miho~Kitamura$^{5},^{\dagger}$}
\author{Koji~Horiba$^5$}
\thanks{Present affiliation: Institute for Advanced Synchrotron Light Source, National Institutes for Quantum and Radiological Science and Technology, 6-6-11 Aoba, Sendai, Miyagi, 980-8579, Japan.}
\author{Teppei~Yoshida$^6$}
\author{Ichiro~Terasaki$^2$}

\affiliation{$^1$Research Institute for Interdisciplinary Science, Okayama University, Okayama 700-8530, Japan} 
\affiliation{$^2$Department of Physics, Nagoya University, Nagoya 464-8602, Japan} 
\affiliation{$^3$Graduate School of Advanced Science and Engineering, Hiroshima University, Higashihiroshima, Hiroshima 739-8521, Japan}
\affiliation{$^4$Hiroshima Synchrotron Radiation Center, Hiroshima University, Higashi-hiroshima 739-0046, Japan }
\affiliation{$^5$Institute of Materials Structure Science, High Energy Accelerator Research Organization (KEK), Tsukuba, Ibaraki 305-0801, Japan}
\affiliation{$^6$Graduate School of Human and Environmental Studies, Kyoto University, Sakyo-ku, Kyoto 606-8501, Japan}

\date{\today}

\begin{abstract}
We report the electronic structure of the thermoelectric semimetal Ta$_2$PdSe$_6$ with a large thermoelectric power factor and giant Peltier conductivity by means of angle-resolved photoemission spectroscopy (ARPES). 
The ARPES spectra reveal the coexistence of a sharp hole band with a light electron mass and a broad electron band with a relatively heavy electron mass, which originate from different quasi-one-dimensional (Q1D) chains in Ta$_2$PdSe$_6$. 
Moreover, the electron band around the Brillouin-zone (BZ) boundary shows a replica structure with respect to the energy originating from plasmonic polarons due to electron-plasmon interactions. 
The different scattering effects and interactions in each atomic chain lead to asymmetric transport lifetimes of carriers: a large Seebeck coefficient can be realized even in a semimetal. 
Our findings pave the way for exploring the thermoelectric materials in previously overlooked semimetals and provide a new platform for low-temperature thermoelectric physics, which has been challenging with semiconductors. 
\end{abstract}

\maketitle

\section{INTRODUCTION}
Semimetal with overlapping conduction and valence bands and coexisting electron and hole carriers have long been studied as the stage for unique physical properties originating from interband interactions. 
In recent years, bulk topological semimetals or Weyl semimetals with the breaking of spatial or time-reversal symmetry have been one of the active areas that have been studied intensively \cite{Shekhar2015, Liang2015}. 
Thermoelectric properties in such semimetals are a fascinating topic. 
Thermoelectric effects like the Seebeck effect or the Peltier effect, which can convert electric thermal energy to electrical energy vice versa, have attracted much attention from the perspective of energy harvesting. 
The thermoelectric conversion efficiency can be expressed as the dimensionless figure of merit $ZT = \frac{S \sigma^2}{\kappa} T$, where $S$ is the Seebeck coefficient, $\sigma$ is the electrical conductivity, $\kappa$ is the thermal conductivity, and $T$ is the absolute temperature. 
In particular, the Seebeck coefficient is given as the energy derivative of the density of states (DOS) for the valence band or conduction band according to Mott’s formula \cite{Mott1936,Cutler1969}. 
Although semimetals possess high electrical conductivity without carrier doping, the magnitudes of the Seebeck coefficients of electron and hole are comparable and cancel each other out, because the shapes of DOS for the valence and conduction bands are symmetric. 
The resultant Seebeck coefficient of semimetals becomes small, and semimetals have long been considered unsuitable for thermoelectric materials. 
Recently, however, a theoretical prediction has been proposed that even semimetals can have large Seebeck coefficients due to the asymmetric DOS shapes of electrons and holes \cite{Markov2018,Markov2019}. 
Furthermore, a giant thermoelectric power factor has been reported in FeSe thin film, gathering attention to thermoelectric semimetals from a theoretical perspective as well \cite{Shimizu2019,Matsubara2023}. 

\begin{figure*}
\includegraphics[width=16cm]{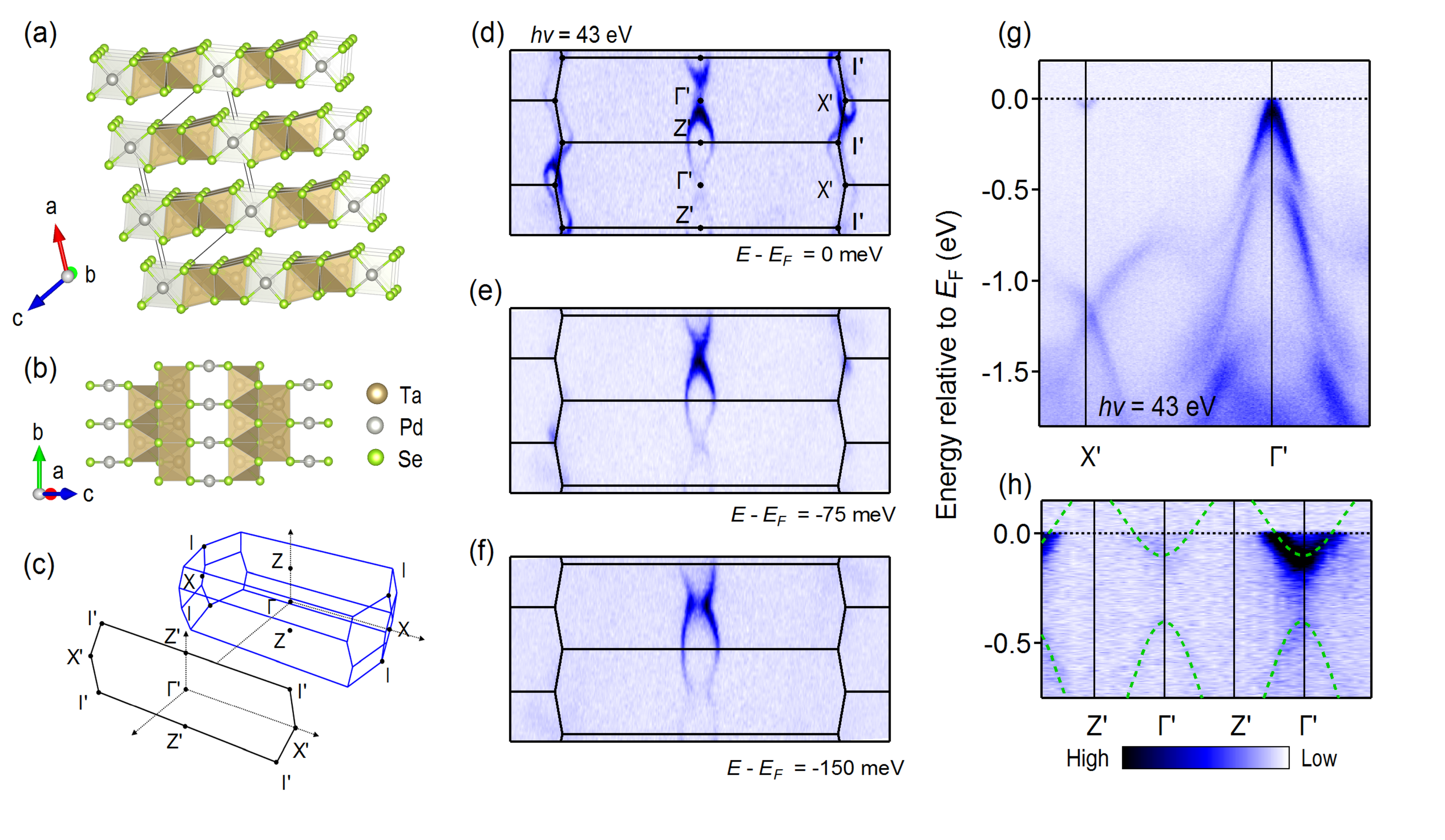}
\caption{(color online) (a), (b) Crystal structure of Ta$_2$PdSe$_6$ visualized by using the software package VESTA \cite{vesta}. 
(c) Brillouin zone (BZ) of Ta$_2$PdSe$_6$. 
The black solid lines indicate the two-dimensional BZ projected on the cleavage plane. 
(d)-(f) FSs and the constant energy contours extracted at $E - E\subm{F}$ = 0, $-75$, and $-150$ meV. 
The ARPES measurements were performed at $h\nu$ = 43 eV. 
Band dispersions along (g) $\Gamma$-X$^\prime$ direction and $\Gamma$-Z direction. 
The green dashed line is the guide to eye.   
The data were collected at $T$ = 20 K. 
} 
\label{f1}
\end{figure*}

An exciting semimetal has been discovered in thermoelectric research: Q1D transition-metal chalcogenide Ta$_2$PdSe$_6$ \cite{Nakano2021,Nakano2021_2,Nakano2022}. 
The crystal structure is lined and stacked in two different chains consisting of the prismatic TaSe$_6$ and the planar PdSe$_4$ (Figs. \ref{f1}(a) - \ref{f1}(c)). 
Ta$_2$PdSe$_6$ coexists with a high electrical conductivity $\sigma$ and a large Seebeck coefficient, which gives a giant Peltier conductivity ($P = S\sigma$). 
This is a measure of electric power generated from a 1 cc thermoelectric material put across a temperature difference of 1 K. 
The maximum PF at low temperature is a remarkably large $\sim 2.4$ mW/cmK$^2$, which is two orders of magnitude larger than bulk thermoelectric materials reported to date. 
Moreover, Ta$_2$PdSe$_6$  also shows giant Peltier conductivity $P = S\sigma$. 
The recorded value $P \sim 100$~A/cmK means that a temperature difference of 1~K on 1~cc sample can generate a current flow of $100$ A. 
However, it remains elusive how such a high Seebeck coefficient at low temperature can be achieved in the semimetal Ta$_2$PdSe$_6$. 
On the basis of two carrier model $S = (\sigma_\mathrm{e} S_\mathrm{e} +\sigma_\mathrm{h} S_\mathrm{h})/(\sigma_\mathrm{e} + \sigma_\mathrm{h})$, the difference of the Seebeck coefficients of electron ($S_\mathrm{e}$) and hole ($S_\mathrm{h}$), or that of their electrical conductivities ($\sigma_\mathrm{h}$ and $\sigma_\mathrm{e}$) can be considered as a contender. 
Clarifying the origin of the high Seebeck coefficient in the semimetal from these contributions would open up a new stage for semimetals as thermoelectric materials and provide guidelines for the material design.

In this work, we report the semimetallic electronic structure of Ta$_2$PdSe$_6$ through direct observation by means of ARPES. 
The Fermi surfaces (FSs) of Ta$_2$PdSe$_6$ consist of two distinctly different components: a sharp hole band with a light effective mass at $Z$ point and a broad electron band with a relatively heavy mass around the Brillouin zone (BZ) boundary originating from the TaSe$_6$ and the PdSe$_4$ chains, respectively. 
A sharpness of the ARPES spectrum represents a long mean free path of quasiparticles (QPs). 
Furthermore, a replica feature of the electron band was observed, suggesting the formation of a plasmonic polaron due to an electron-plasmon coupling. 
The polaronic feature, only observed in the electron band, would cause the difference in the transport scattering time between the hole and electron carriers.  
These asymmetric electronic structures between the hole and electron bands would be responsible for the high Seebeck coefficient of Ta$_2$PdSe$_6$. 
Our spectroscopic findings reveal the essential role of the hole and electron FSs for the high Seebeck coefficient of Ta$_2$PdSe$_6$ and provide realistic guidelines for the material design for highly effective thermoelectric semimetals.

\section{EXPERIMENTAL SETUP}
A high-quality single crystal of Ta$_2$PdSe$_6$ was grown by the self-flux method \cite{Nakano2021}. 
ARPES measurements were carried out at BL-28A of Photon Factory with a Scienta Omicron DA30 electron analyzer \cite{Kitamura2022} and at BL-9A of HiSOR with a Scienta Omicron R4000 (SPECS PHOIBOS 150)  electron analyzer. 
The incident photon energy was set to $h\nu$ = 81, 43, and 21 eV with the circular polarized light. 
The total energy resolution was set at 20 meV for $h\nu$ = 81 eV, 11 meV for $h\nu$ = 43 eV, and 10 meV for $h\nu$ = 21 eV. 
The measurement temperature was $T$ = 20 K, and the main chamber was maintained in an ultra-high vacuum of higher than 2.1$\times$10$^{-10}$ Torr. 
To obtain clean surfaces for the photoemission measurement, the samples were cleaved $in$ $situ$ at 20 K.  
The binding energy was calibrated by using the Fermi edge of the gold reference.

\section{RESULTS AND DISCUSSION}
\begin{figure*}
\includegraphics[width=18cm]{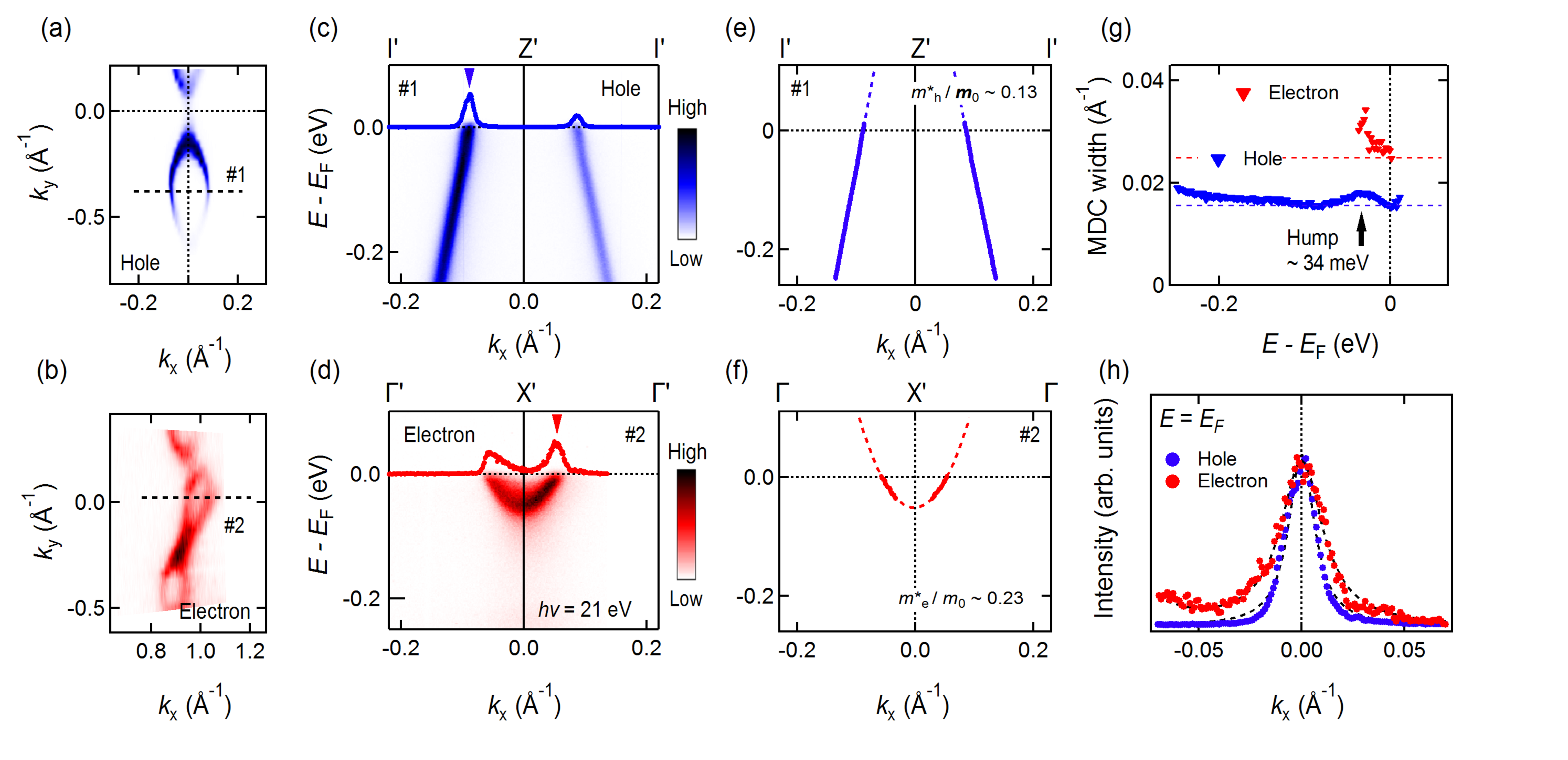}
\caption{(color online) (a),(b) FSs consisting of hole and electron bands. 
(c),(d) Band dispersions along $\Gamma$-I$^\prime$ and X$^\prime$-$\Gamma$ directions taken from the cuts \#1 and \#2 in (a) and (b). 
The ARPES data were collected at $h\nu$ = 21 eV. 
(e) Hole and (d) electron bands extracted from the Lorentz fitting of MDCs in (c) and (d). 
(g) MDC widths for the electron and hole bands as a function of energy. 
The arrow indicates the energy position of the kink structure. 
(h) MDCs at $E\subm{F}$ for the hole and electron bands taken from (c) and (d). 
} 
\label{f2}
\end{figure*}

Figure \ref{f1}(d) shows the FSs of Ta$_2$PdSe$_6$ taken at $h\nu$ = 43 eV. 
The parallel wavy intensities were observed along the $\Gamma$-Z and X$^\prime$-I$^\prime$ directions. 
In the constant energy contours of Figs. \ref{f1}(e) and \ref{f1}(f), the area of intensities around $\Gamma$-Z expands in going from $E\subm{F}$ to $-150$ meV. 
On the other hand, the intensities stick to the BZ boundary (X$^\prime$-I$^\prime$ line) vanish at $E-E\subm{F}$ = $-150$ meV. 
Indeed, the hole-like band around $\Gamma$ point and the electron band around X$^\prime$ point were identified from the band dispersion in Fig. \ref{f1}(g). 
Although the top of the hole band at $\Gamma$ point is located below $E\subm{F}$, the large hole pockets are formed around Z point. 
Here, we underline that the FS around the $Z$ point (the BZ boundary) is the warped cylindrical hole pocket (the two-dimensional electron pocket). 
The photon-energy dependence is included in the Supplemental Materials \cite{Suppl}. 
Although the surface and bulk separation is difficult in the low-dimensional system, the observed FSs and band dispersions agree well with the band-structure calculation considering a mBJ potential with the spin-orbit interaction \cite{Yang2022, Nakano2025, Suppl} or a Hubbard $U$ \cite{Wang2024}. 
We consider that our results reflect the bulk electronic structure, rather than a topological surface state reported in a recent ARPES study \cite{Yang2025}. 
Our results reveal the overlap of the electron and hole bands: the semimetallic state is realized in Ta$_2$PdSe$_6$.

To address more details of the semimetallic state directly related to the thermoelectric properties of Ta$_2$PdSe$_6$, we have extracted the hole and electron bands forming hole and electron pockets. 
Figures \ref{f2}(c) and \ref{f2}(d) show the ARPES intensities of the hole and electron bands corresponding to the cuts \#1 and \#2 in Figs. \ref{f2}(a) and \ref{f2}(b), respectively. 
The band dispersions taken from the Lorentz fitting of the momentum distribution curves (MDCs) are displayed in Figs. \ref{f2}(e) and \ref{f2}(f). 
The effective hole (electron) mass was estimated to be $m^*_h/m_0 \sim 0.13$ ($m^*_e/m_0 \sim 0.23$) from the Fermi velocities of the band dispersions in Figs. \ref{f2}(e) and \ref{f2}(f). 
Here, the Fermi velocity of the hole (electron) was $v\subm{F}^h \sim 8.0 \times 10^{5}$ m/s ($v\subm{F}^e \sim 2.9 \times 10^{5}$ m/s). 
The effective mass of the hole band is light, which is almost 1.8 times smaller than that of the electron band ($m^*_h/m^*_e \approx 0.56$). 
Moreover, we have plotted the MDC width $\Delta k$ related to the QP lifetime as a function of energy in Fig. \ref{f2}(g). 
The width of the hole band is narrower than that of the electron band in the whole energy region. 
The direct comparison of the MDC at $E\subm{F}$ is displayed in Fig. \ref{f2} (h). 
The width of the hole band is $\Delta k^h\subm{F}$ = 0.016 \AA$^{-1}$, and that of the electron band is $\Delta k^e\subm{F}$ = 0.025 \AA$^{-1}$. 
This means that the mean free path $l = 1/\Delta k$ is distinct between holes and electrons ($l^h>l^e$). 
From the QP lifetime follows $\tau = 1/v\subm{F} \Delta k = l/v\subm{F}$, the hole and electron mobility $\mu_h$ and $\mu_e$ can be obtained by using the QP lifetime $\tau$ and the effective mass $m^*$ for hole and electron ($\mu = e\tau/m^*$). 
Interestingly, the hole and electron mobility ratio is estimated to be $\mu_e/\mu_h \approx 0.84$, while the mean free path $l^h$ of holes is longer than that $l^e$ of electrons. 
This is because the QP lifetime ratio $\tau_e / \tau_h = 1.48$ compensates for the effective mass ratio $m^*_h / m^*_e  = 0.56$. 
These results appear to contradict that ratio $\mu_e/\mu_h \approx 0$ in the previous transport studies \cite{Nakano2021, Nakano2021_2, Nakano2022, Kato2024, Nakano2024_2, Yang2025, Nakano2025}. 
We speculate that this discrepancy may originate from the dimensionalities of electron and hole carriers within the two-carrier model.

\begin{figure}
\includegraphics[width=9cm]{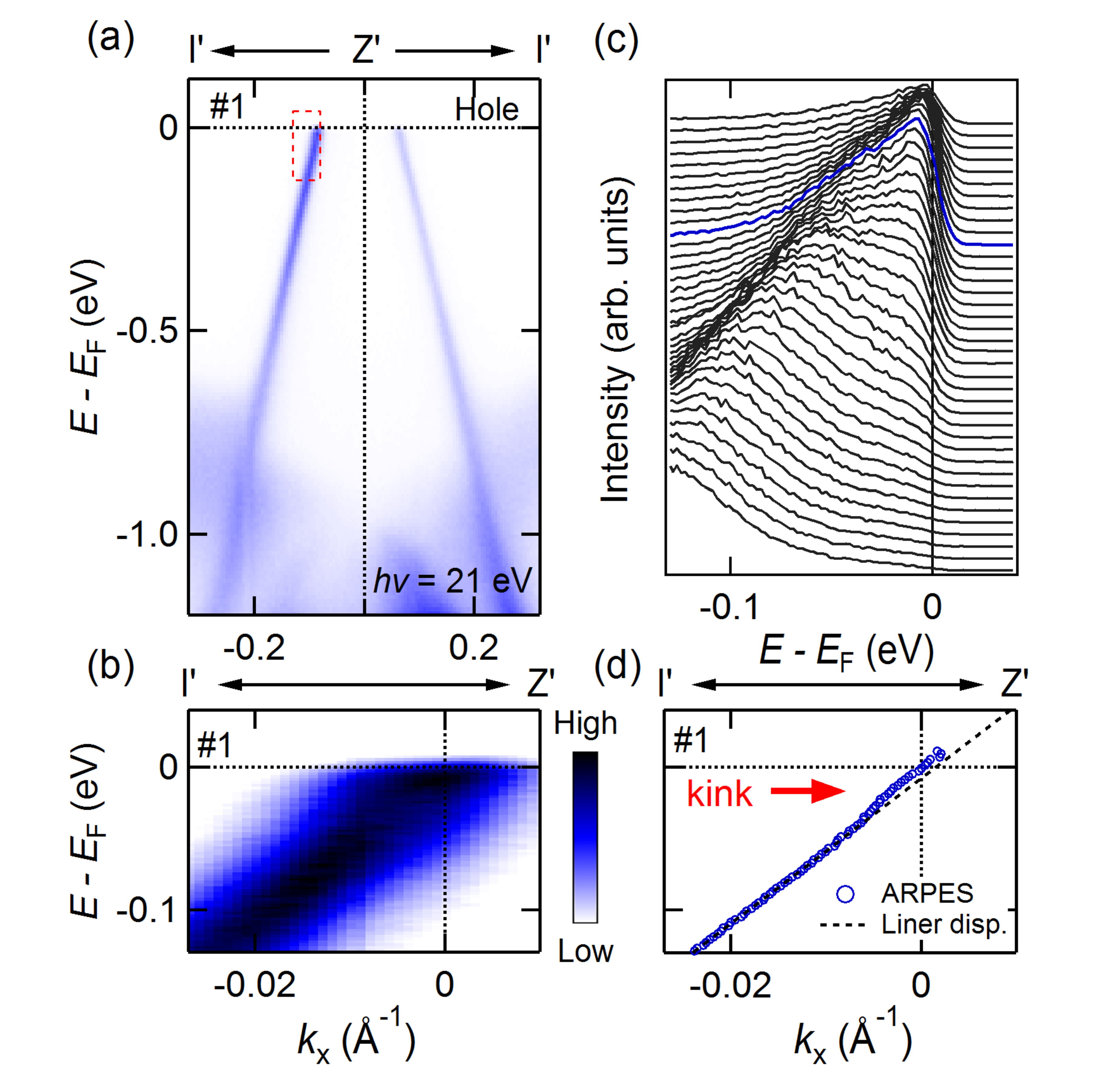}
\caption{(color online) (a) Hole-band dispersion around Z$^\prime$ measured at $h\nu$ = 21 eV and (b) its enlarged view of (a). 
(c) EDCs of (b). 
(d) Hole band extracted from the Lorentz fitting of MDCs in (b). 
The momentum is rescaled by $k\subm{F}$ of the hole band in (c) and (d). 
The red arrow is the position of the kink structure, and the dashed line indicates the fitted result of the data at lower energies than the kink energy. 
} 
\label{f3}
\end{figure}

We emphasize that the large difference of $l$ is rare and indicates the qualitatively different electronic structures between the hole and the electron, although the line width of the ARPES spectra indicates the QP lifetime, and the transport scattering time $\tau$ originates mainly from the backscattering factor of the Boltzmann transport equation. 
The ARPES intensity of the hole band is displayed in Fig. \ref{f3}(a), corresponding to the cut \#1 of Fig. \ref{f2}(a). 
Figures \ref{f3}(b) and \ref{f3}(c) show the magnified view and the energy distribution curves (EDCs) of the rectangular area in Fig. \ref{f3}(a). 
We have deduced the hole-band dispersion around $Z^\prime$ point by the Lorentz fitting for the MDCs, 
The resultant hole band (open circles) deviates from the linear dispersion (dashed line), indicating the kink structure at $\sim-15$ meV due to some electron-boson coupling as shown in Fig. \ref{f3}(d). 
This feature can also be seen as the hump structure around $-34$ meV observed in the energy dependence of the MDC width for the hole band as shown in Fig. \ref{f2}(g).
Here, the detection of such a weak kink structure reflects the high energy and wavenumber resolution in the present study. 
Noted that the Debye energy $k\subm{B}\Theta\subm{D} = 15.3$ meV ($\Theta\subm{D} = 174$ K) agrees with the energy scale of the kink energy \cite{Kato2024}. 
Thus, we speculate that the observed kink structure originates from the electron-phonon interaction.

\begin{figure}
\includegraphics[width=9cm]{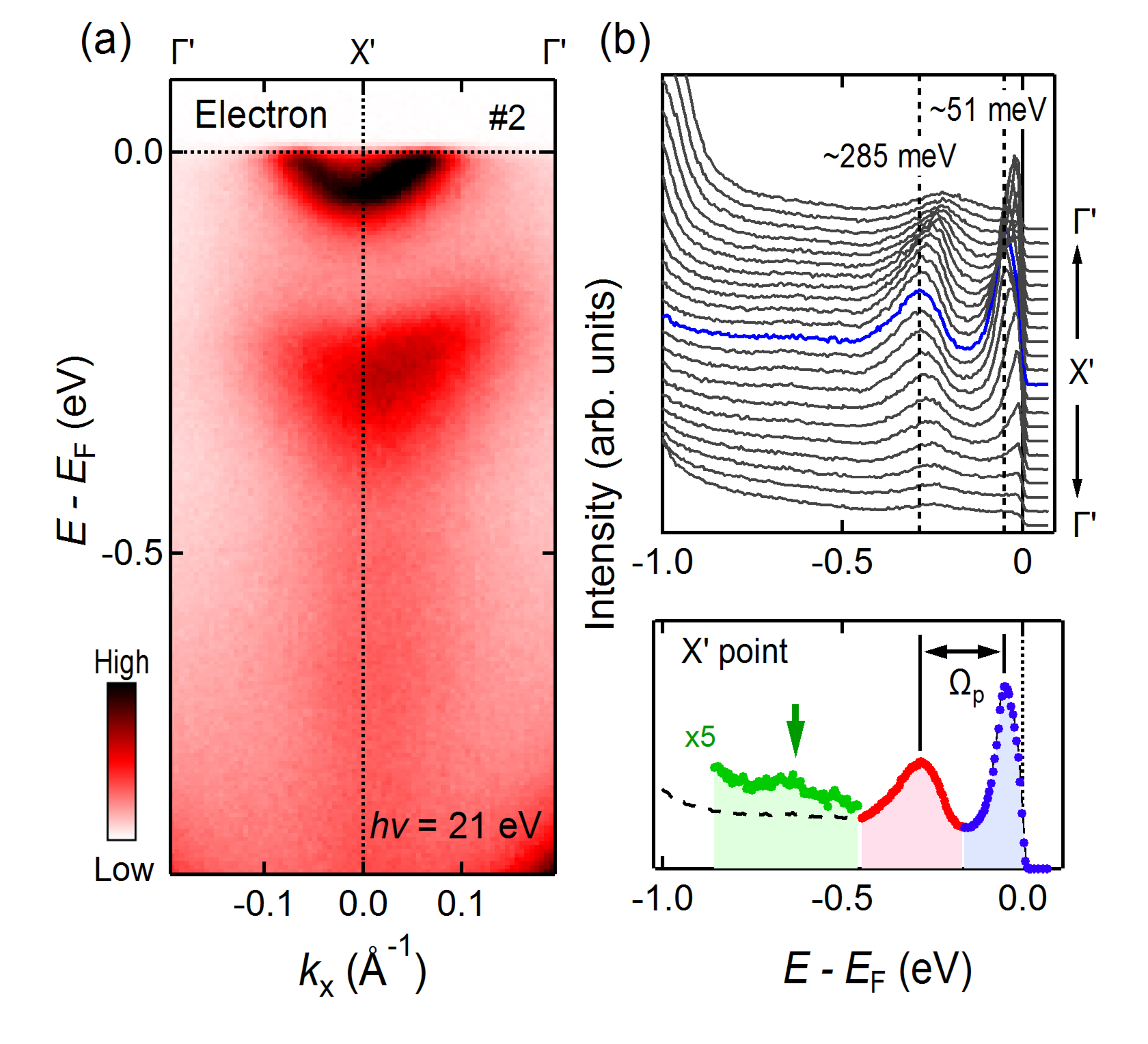}
\caption{(color online) (a) Electron-band dispersion measured at $h\nu$ = 21 eV and (b) its EDCs of (a). 
The lower panel of (b) indicates the EDC at $\Gamma$ point taken from (a). 
The intensity from $E-E\subm{F}$ = $-0.48$ eV to $-0.87$ eV is multiplied by 5 to visualize the second-order replica intensity. 
The black dotted line indicates the EDC at $\Gamma$ point as it is. 
} 
\label{f4}
\end{figure}

Figure \ref{f4}(a) shows the ARPES intensity of the electron band, and the strong intensity consisting of the electron FS was identified. 
Furthermore, the replica band can be identified below the bottom of the electron band. 
To visualize more, the EDCs are displayed in Fig. \ref{f4}(b). 
The bottom of the electron (replica) band is located around $-51$ meV ($-285$ meV). 
Furthermore, the weak intensity was observed around $-700$ meV as shown in the lower panel of Fig. \ref{f4}(b). 
The observed replica structure shows a larger modulation than the kink structure observed in the hole band in Figs. \ref{f2}(g) and \ref{f3}(c). 
The observed replica band is a characteristic feature observed spectroscopically and suggests the existence of polarons. 
Here, we consider that the polaron feature originates from the intrinsic electron-boson coupling in the initial state \cite{Faeth2021}, not for the extrinsic final-state energy loss feature, because the intensity of the polaron feature does not depend on the photon energy \cite{Li2018}, as shown in the Supplemental Materials \cite{Suppl}. 
Interestingly, this polaron structure, which is absent in the hole band [Fig. \ref{f3}], would provide strong scattering for carriers on the electron FS and affect the scattering time of electrons. 
We speculate that the polaron structure observed only on the electron band causes significantly different mobility than the hole carriers. 
As supporting evidence, the previous band calculation reported that the hole FS originates from the Se $4d$ orbitals of the PdSe$_4$ chains since the Pd $4d$ orbital is almost fully occupied \cite{Nakano2021_2}. 
On the other hand, the electron FS is derived from the Ta $5d$ orbital strongly hybridized with the Se $4d$ orbitals of the TaSe$_6$ chain \cite{Nakano2021_2, Nakano2025}.  
Thus, the unique crystal structure of Ta$_2$PdSe$_6$, in which each chain forms the electron and hole FSs, respectively, only brings the polaron feature in the electron FS.

Finally, we discuss the origin of the observed polaron structure. 
The energy separation of the polaron structure can be estimated to be $\approx$ 234 meV from the peak position in EDC at X$^\prime$ point [Fig. \ref{f4}(b)], reflecting the energy scale of the polaron energy $\hbar \Omega\subm{p}$. 
This value is unexpectedly larger than that in transition-metal oxides such as TiO$_2$, SrTiO$_3$, and mono-layer FeSe on SrTiO$_3$, where the polaron can be associated with a longitudinal optical (LO) phonon and the optical phonon branch \cite{Moser2013, Chen2015, Wang2016, Yukawa2016, Cancellieri2016, Lee2014, Zhang2017, Liu2021}. 
Actually, the highest value of the phonon mode of Ta$_2$PdSe$_6$ is about 33 meV from the calculated phonon dispersion \cite{Suppl}, which is lower energy than that for the oxides as well as the observed polaron energy. 
These facts show that the simple polaron picture derived from the electron-phonon interaction cannot explain the large energy separation observed.

To understand the large value of this system, we have applied the plasmonic polaron model \cite{Caruso2015, Caruso2018, Riley2018,Caruso2021,Ulstrup2024}. 
The plasmonic polaron forms when the conductive electron interacts with the charge-density oscillation known as a plasmon. 
The energy scale of the plasmonic polaron is proportional to the plasma frequency: $\omega\subm{p} = (n e^2/ m^* \epsilon_0 \epsilon_\infty )^{1/2}$, where $n$ is the carrier density, $m^*$ is the effective mass, $\epsilon_0$ is the dielectric permittivity of free space, and $\epsilon_\infty$ is the dielectric constant. 
We set the parameters $n = 7.0 \times 10^{20}$ cm$^{-3}$ \cite{Nakano2021_2}, $m^*_e = 0.9 m_0$ \cite{Nakano2021_2, Nakano2022} from the transport study on Ta$_2$PdSe$_6$. 
On the other hand, little is known about the dielectric permittivity $\epsilon_\infty $ of Ta$_2$PdSe$_6$. 
By assuming $\epsilon_\infty \approx 19.4$, the estimated plasma frequency $\hbar\Omega\subm{p}$ corresponds to the observed energy separation. 
This permittivity is very close to $\epsilon_\infty = 22$ of Ta$_2$NiSe$_5$, which has a similar crystal structure of Ta$_2$PdSe$_6$, showing the semiconducting behavior due to the excitonic insulator \cite{Nakano2019}. 
Even in the three-dimensional Dirac semimetal Cd$_3$As$_2$, the permitivity is $\epsilon_\infty = 20$ - $44$ \cite{Zivitz1974, Throckmorton2015}. 
Moreover, the plasma frequency of a Q1D material TaSe$_3$ has been reported to be 680 meV along chains and 420 meV perpendicular to chains \cite{Geserich1986}. 
The resultant value agrees with the energy separation of the polaron structure observed in Fig. \ref{f4}(b). 
Therefore, it is reasonable to attribute the replica feature observed only in the electron band to the plasmonic polarons.

\section{CONCLUSION}
In summary, we have studied the electronic structure of the thermoelectric semimetal Ta$_2$PdSe$_6$ with the large thermoelectric power factor and the giant Peltier conductivity using ARPES. 
The hole and electron FSs were observed in the BZ center and boundary, respectively. 
The hole band overlaps the electron band, indicating the semimetallic state in Ta$_2$PdSe$_6$. 
Each band is derived from the TaSe$_6$ and PdSe$_4$ chains. 
The sharp hole band with the light effective mass coexists with the broad electron band with the relatively low effective mass, showing the asymmetric electronic structure of the hole and electron bands. 
Moreover, the replica structure was observed in the electron band, which suggests the signs of the polaron due to the electron-plasmon interaction. 
The polaron feature further promotes the electron-hole asymmetry through the enhancement of the strong backscattering in the electron band, which would provide a large Seebeck coefficient even in semimetals. 
Our spectroscopic findings provide realistic guidelines using the QP anomaly for the material design of highly effective thermoelectric semimetals. 

\section*{ACKNOWLEDGMENTS}
The authors would like to thank H.~Fukuyama and M.~Matsubara for fruitful discussions, T.~Ishida for the experimental support, A.~Honma, S.~Souma, K.~Ozawa, and T.~Sato for the technical assistance of BL-28A at Photon Factory. 
This work was supported by Japan Society for the Promotion of Science (JSPS) Grants-in-Aid for Scientific Research (KAKENHI Nos. 17H06136, 21K13878, 21K13882, and 23K13059). 
The synchrotron radiation experiment was performed with the approvals of Photon Factory (Proposal Nos. 2018S2-001, 2019G122, 2021G101, 2021S2-001, 2022G077, and 2024G081) and HSRC (Proposal No. 21AG030, No. 21BG022, and No. 22BG010).

\section*{DATA AVAILABILITY}
The data that support the findings of this study are available from the corresponding author upon reasonable request.

\end{document}